\documentclass[
  ,final            
  ]
  {aipproc}

\usepackage{bm}

\layoutstyle{8x11single}

\def\ecc{\varepsilon}
\def\es{(\eta/s)_\mathrm{QGP}}
\def\esb{(\eta/s)_\mathbf{QGP}}
\newcommand{\U}{{\small UrQMD}\,}
\newcommand{\V}{{\small VISHNU}\,}
\newcommand{\VISH}{{\small VISH2{+}1}\,}

\def\La{\langle}
\def\Ra{\rangle}

\newcommand{\dNdy}{dN_\mathrm{ch}/dy}

\newcommand{\eq}{{\,=\,}}

\newcommand{\Tc}{T_\mathrm{c}}
\newcommand{\Tchem}{T_\mathrm{chem}}

\begin{document}

\title{The Viscosity of Quark-Gluon Plasma at RHIC and the LHC}


\author{\underbar{Ulrich Heinz}}{
  address={Department of Physics, The Ohio State University, Columbus, Ohio 43026, USA}
}
\author{Chun Shen}{
  address={Department of Physics, The Ohio State University, Columbus, Ohio 43026, USA}
}
\author{Huichao Song}{
  address={Lawrence Berkeley National Laboratory, Berkeley, California 94720, USA}
}	

\begin{abstract}
The specific shear viscosity $\es$ of quark-gluon plasma (QGP) can 
be extracted from elliptic flow data in heavy-ion collisions by comparing 
them with the dynamical model {\footnotesize VISHNU} which couples a viscous fluid dynamic 
description of the QGP with a microscopic kinetic description of the late hadronic 
rescattering and freeze-out stage. A robust method for fixing $\es$ from the 
collision centrality dependence of the eccentricity-scaled charged hadron elliptic flow is presented. The 
systematic uncertainties associated with this extraction method are discussed, with specific attention to our presently restricted knowledge of initial conditions. With the $\es$ extracted
in this way, {\footnotesize  VISHNU} yields an excellent description of all soft-hadron data 
from Au+Au collisions at top RHIC energy. Extrapolations to Pb+Pb collisions at the LHC, 
using both a purely hydrodynamic approach and {\footnotesize VISHNU}, are presented 
and compared with recent experimental results from the ALICE Collaboration. The LHC 
data are again well described by {\footnotesize VISHNU}, with the same $\es$ 
value as at RHIC energies. 
\end{abstract}

\maketitle


\section{How to measure $\bm{\esb}$}
\label{sec1}

Relativistic heavy-ion collisions create spatially deformed fireballs
of hot, dense matter -- in both non-central and (due to event-by-event
shape fluctuations) central collisions. Hydrodynamics converts this 
initial spatial deformation into final state momentum anisotropies. Viscosity 
degrades the conversion efficiency 
$\ecc_x\eq\frac{\La\!\La y^2{-}x^2\Ra\!\Ra}{\La\!\La y^2{+}x^2\Ra\!\Ra}
\to 
\ecc_p\eq\frac{\La T^{xx}{-}T^{yy}\Ra}{\La T^{xx}{+}T^{yy}\Ra}$ 
of the fluid ($x$ and $y$ are the directions transverse to the beam direction 
$z$); for given initial fireball ellipticity $\ecc_x$, the viscous 
suppression of the dynamically generated total momentum anisotropy 
$\ecc_p$ is monotonically related to the specific shear viscosity $\eta/s$. 
The observable most directly related to $\ecc_p$ is the total charged 
hadron elliptic flow $v_2^\mathrm{ch}$ \cite{Heinz:2005zg}. Its 
distribution in $p_T$ depends on the chemical composition and 
$p_T$-spectra of the various hadron species; the latter evolve in
the hadronic stage due to continuously increasing radial flow (and 
so does $v_2(p_T)$), even if (as expected at top LHC energy 
\cite{Hirano:2007xd}) $\ecc_p$ fully saturates in the QGP phase. When 
(as it happens at RHIC energies) $\ecc_p$ does not reach saturation before 
hadronization, dissipative hadronic dynamics \cite{Hirano:2005xf} affects 
not only the distribution of $\ecc_p$ over hadron species and $p_T$, but 
even the final value of $\ecc_p$ itself, and thus $v_2^\mathrm{ch}$ from 
which we want to extract $\eta/s$. To isolate the QGP viscosity $\es$ we 
therefore need a hybrid code that couples viscous hydrodynamics of the 
QGP to a realistic model of the late hadronic stage, such as \U
\cite{Bass:1998ca}, that describes its dynamics microscopically. 
\V \cite{Song:2010aq}, a hybrid of \VISH ({\bf V}iscous {\bf I}srael-{\bf S}tewart 
{\bf H}ydrodynamics in {\bf 2+1} dimensions \cite{Song:2007fn}) and 
{\small UrQMD}, is such a code.

\section{$\bm{\esb}$ at RHIC}
\label{sec2}

\begin{figure}[htb]
 \includegraphics[width=0.35\linewidth,clip=,angle=0]{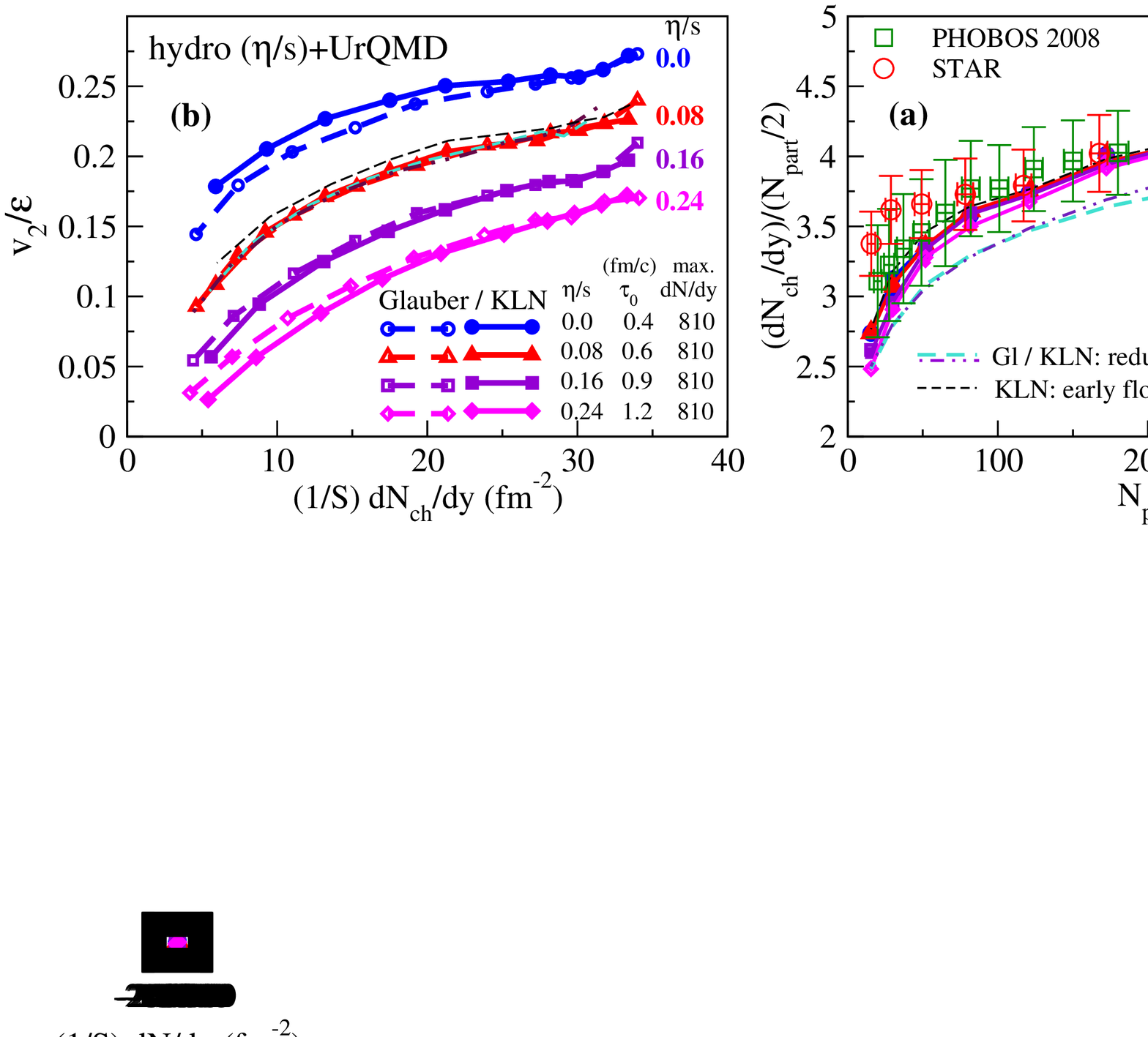}  
 \includegraphics[width=0.635\linewidth,clip=,angle=0]{fig1b.eps}  
\caption{\label{F1} (Color online) Centrality dependence of 
eccentricity-scaled elliptic flow \cite{Song:2010mg}.
}
\end{figure}

The left panel in Fig.~\ref{F1} shows that such an approach yields a
universal dependence of the ellipticity-scaled total charged hadron 
elliptic flow, $v_2^\mathrm{ch}/\ecc_x$, on the charged hadron multiplicity 
density per overlap area, $(1/S)(\dNdy)$, that depends only on 
$\es$ but not on the details of the initial state model that provides $\ecc_x$ 
and $S$ \cite{Song:2010mg}. Pre-equilibrium flow and bulk viscous effects 
on these curves are small \cite{Song:2010mg}. 

The QGP viscosity can be extracted from experimental $v_2^\mathrm{ch}$ data 
by comparing them with these universal curves. The right panels of Fig.~\ref{F1} 
show this for MC-Glauber and MC-KLN initial state models (please see 
\cite{Song:2010mg} and references therein for a description of these models). In 
both cases the slope of the data \cite{Ollitrault:2009ie} is correctly reproduced; this
is not the case for ideal nor for viscous hydrodynamics with constant $\eta/s$. Due 
to the ${\sim}20\%$ larger ellipticity of the MC-KLN fireballs, the magnitude of 
$v_\mathrm{2,exp}^\mathrm{ch}/\ecc_x$ differs between the two models. 
Consequently, the value of $(\eta/s)_\mathrm{QGP}$ extracted from this 
comparison changes by more than a factor 2 between them. Relative to the 
initial fireball ellipticity all other model uncertainties are negligible. Without 
constraining $\ecc_x$ more precisely, $\es$ cannot be determined to better 
than a factor 2 from elliptic flow data alone, irrespective of any other model 
improvements.\footnote{It has been suggested 
  \cite{Shen:2011zc,Qiu:2011fi,:2011vk,Adare:2011tg} that the ambiguity 
  between the MC-Glauber and MC-KLN ellipticities which lies at the origin 
  of this uncertainty can be resolved by simultaneously analyzing elliptic and triangular 
  flow, $v_2$ and $v_3$.} 
Taking the MC-Glauber and MC-KLN models to represent 
a reasonable range of initial ellipticities, Fig.~\ref{F1} gives $1{\,<\,}4\pi\es{\,<\,}2.5$ for 
temperatures $\Tc{\,<\,}T{\,<\,}2\Tc$ probed at RHIC. 

All calculations in Fig.~\ref{F1} and following below were done in "single-shot" mode, 
where the ensemble of fluctuating Monte Carlo initial states was first averaged in the 
participant plane \cite{Song:2010mg} to obtain a smooth average initial density profile 
and then evolved just once through the hydrodynamic stage. Event-by-event evolution 
of each fluctuating initial state separately and performing the ensemble average only 
at the end may produce somewhat less elliptic flow and thus slightly reduce the $\es$ 
values extracted from comparison with the data \cite{Qiu:2011iv}. The magnitude of this 
reduction depends on $\es$ \cite{Qiu:2011fi} but is not expected to exceed 
(0.2-0.3)/4$\pi$ \cite{Qiu:2011iv}.

\begin{figure}[htb]
 \includegraphics[width=0.47\linewidth,clip=,angle=0]{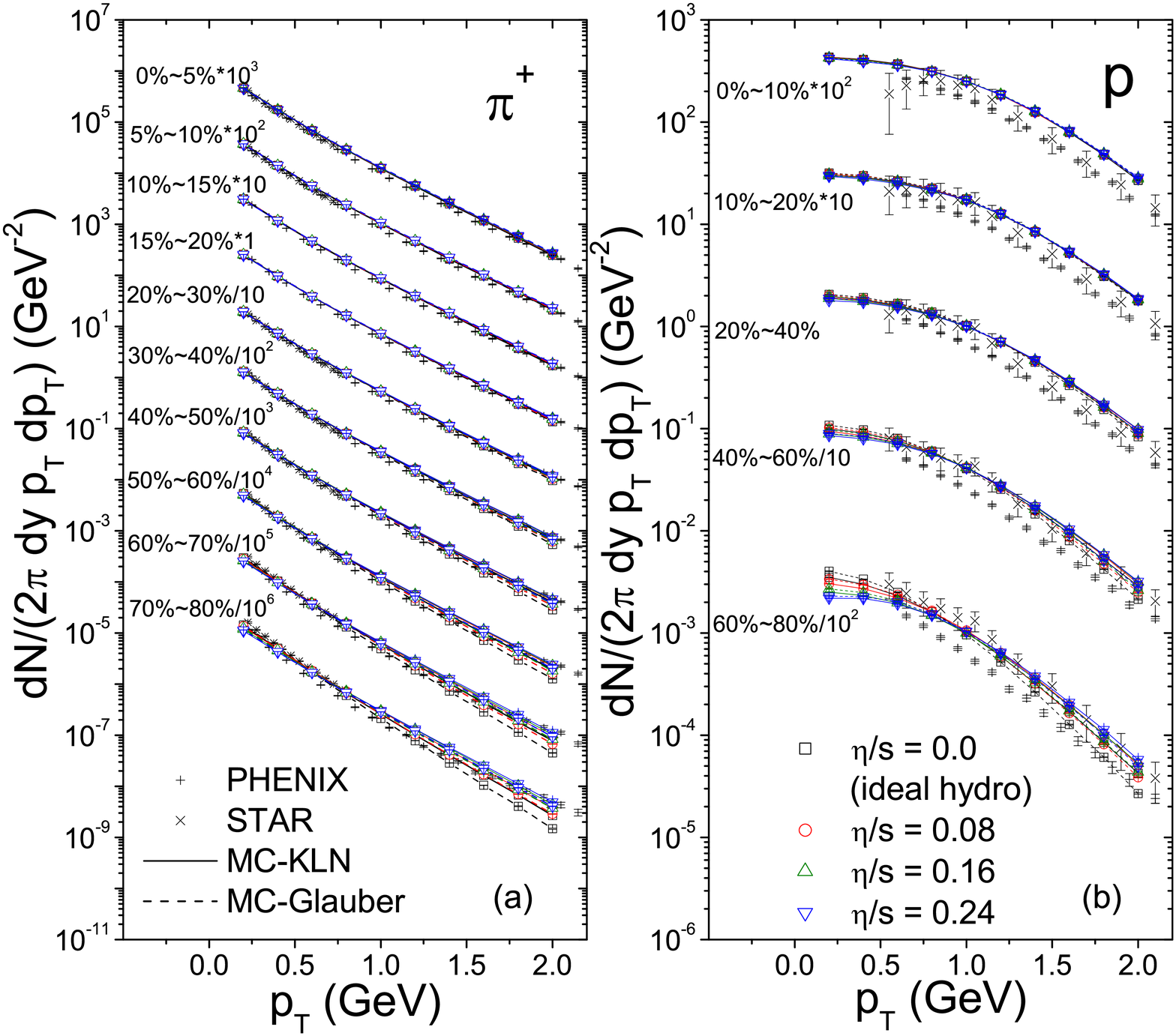}  
 \includegraphics[width=0.53\linewidth,clip=,angle=0]{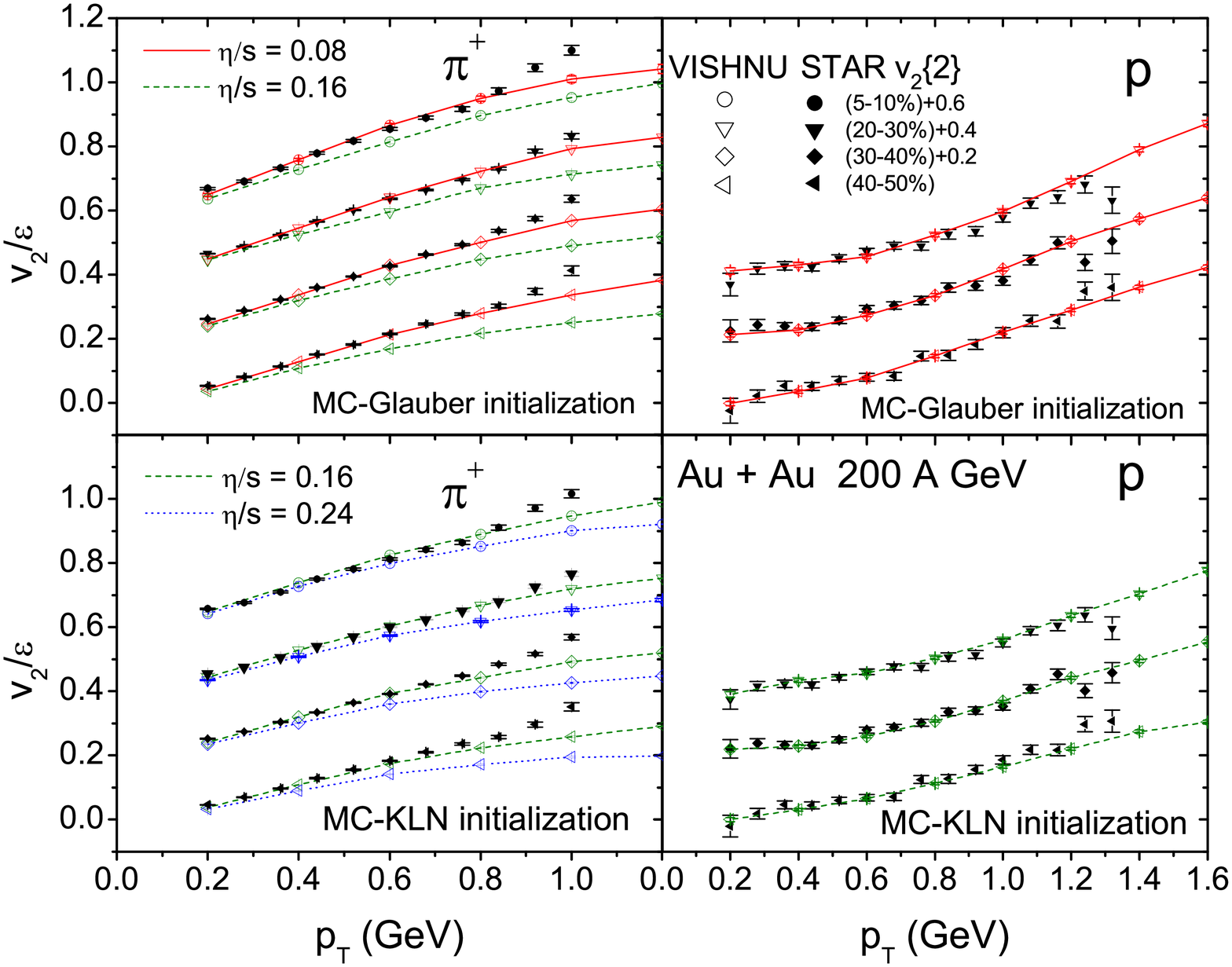}  
 \caption{\label{F2} (Color online) Transverse momentum spectra (left panel) 
   and differential elliptic flow $v_2(p_T)$ (right panel) for identified pions and 
   protons from 200\,$A$\,GeV Au+Au collisions at RHIC for different centralities.
   Experimental data are from STAR and PHENIX, theoretical lines from {\footnotesize
   VISHNU} (see \cite{Song:2011hk} for details and references). 
   After appropriate adjustment of initial conditions \cite{Song:2011hk}
   the $p_T$-spectra are seen to be insensitive to the QGP viscosity whereas
   the elliptic flow depends strongly on it. For MC-Glauber initial conditions 
   (upper right panels) $\es\eq0.08$ works well, for MC-KLN (bottom right
   panels) $\es\eq0.16$ works well for all collision centralities. In each case,
   changing $\es$ by 0.08 destroys the agreement between theory and data. 
   }
\end{figure}

\V with $\es\eq\frac{1} {4\pi}$ for MC-Glauber and $\frac{2}{4\pi}$ for MC-KLN provides an 
excellent description of all aspects of soft ($p_T{\,<\,}1.5$\,GeV) hadron production 
($p_T$-spectra and differential $v_2(p_T)$ for all charged hadrons together as well as 
for individual identified species) in 200\,$A$\,GeV Au+Au collisions at 
all but the most peripheral collision centralities \cite{Song:2011hk}. As an example we show
in Fig.~\ref{F2} $p_T$-spectra and differential elliptic flow for identified pions and
protons (resonance decay contributions are included). Such a level of theoretical control is unprecedented.\footnote{We note that the purely hydrodynamic model VISH2{+}1
  does almost equally well, with $\es\eq0.2$ for MC-KLN initial conditions 
  \cite{Shen:2011eg}, except for the centrality dependence of the differential 
  elliptic flow $v_2(p_T)$ for protons. We will see a similar failure of  VISH2{+}1 for
  Pb+Pb collisions at the LHC further below. The main difference to VISHNU is that, in 
  order to generate enough radial flow at freeze-out, VISH2{+}1 must be started earlier 
  ($\tau_0\eq0.6$ instead of 1.05\,fm/$c$) because it lacks the highly dissipative
  hadronic phase that generates additional radial flow in VISHNU at late times (in the 
  VISH2{+}1 simulations $\eta/s$ is held constant at 0.2 until hadronic freeze-out). The 
  variation with collision centrality of the final balance between radial and elliptic flow 
  turns out to be correct in VISHNU (where more of the radial flow develops later) but 
  incorrect in VISH2{+}1 (where more of it is created early).}

\section{$\bm{\esb}$ at the LHC}
\label{sec3}

\begin{figure}[h!]
  \includegraphics[width=0.45\linewidth,height=5.5cm,clip=,angle=0]%
                  {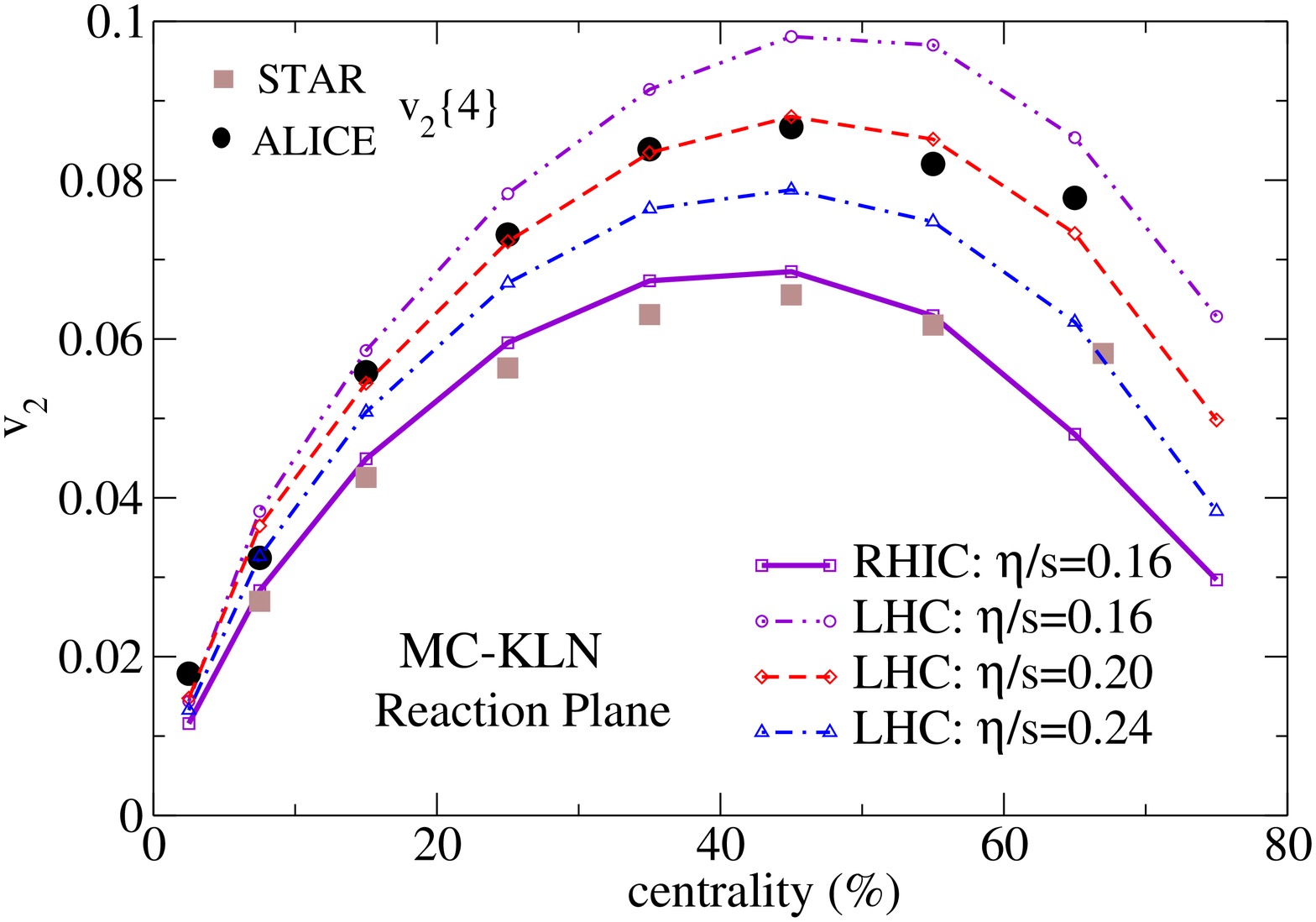}
  \includegraphics[width=0.48\linewidth,height=5.58cm,clip=,angle=0]{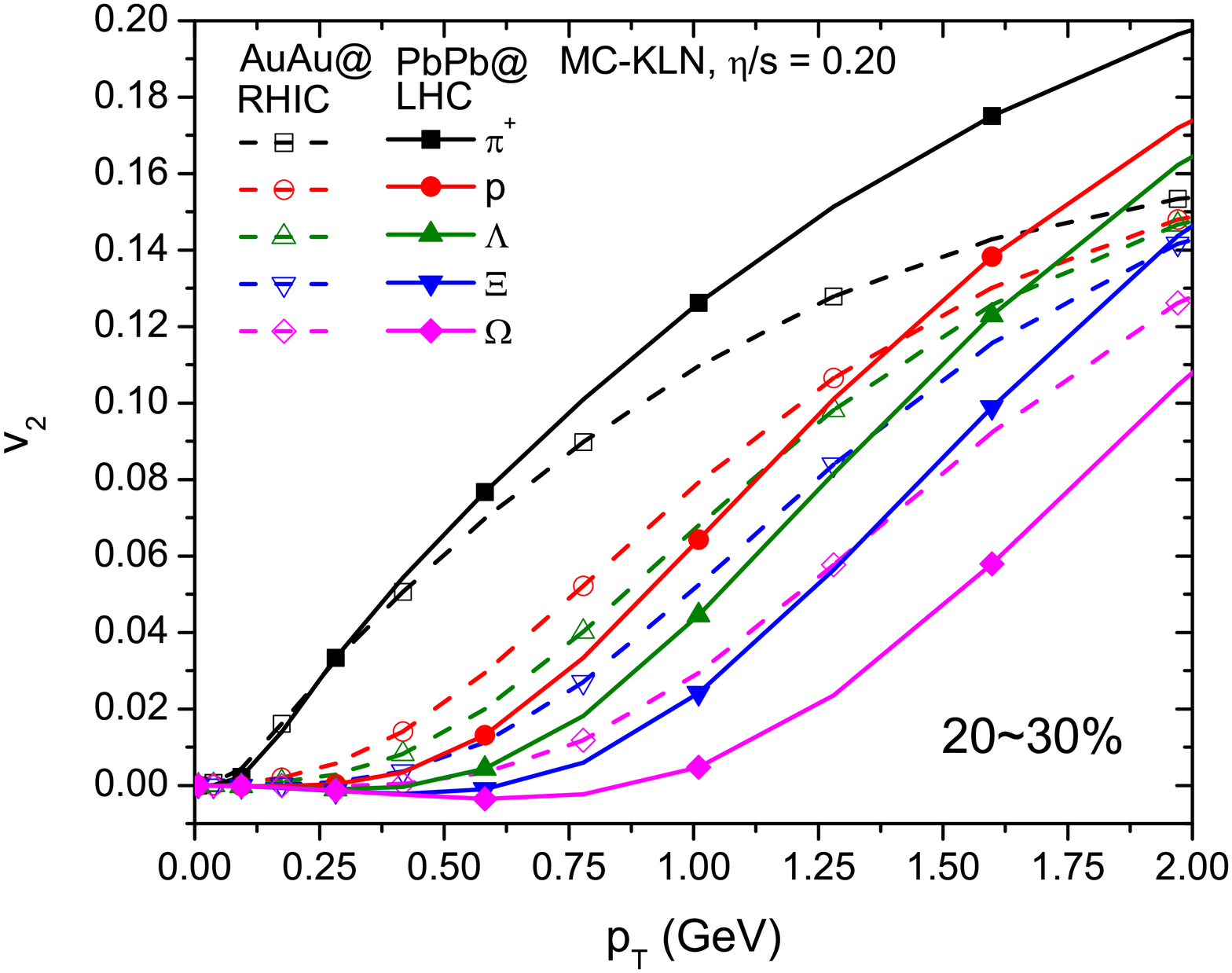}  
\caption{\label{F3} (Color online) Total charged hadron elliptic flow as
function of centrality (VISHNU, left \cite{Song:2011qa}) and 
differential elliptic flow for identified hadrons for 20-30\% centrality 
(VISH2{+}1, right \cite{Shen:2011eg}) for 200\,$A$\,GeV Au+Au collisions 
at RHIC and 2.76\,$A$\,TeV Pb+Pb collisions at the LHC. Experimental data 
are from \cite{Aamodt:2010pa}.
        }
\end{figure}

The successful comprehensive fit of soft hadron spectra and elliptic flow in Au+Au 
collisions at RHIC shown in Fig.~\ref{F2} and elaborated on in more detail in 
Refs.~\cite{Song:2011hk,Shen:2011eg} allows for tightly constrained LHC predictions.
Fig.~\ref{F3} shows such predictions for both pure viscous hydrodynamics \VISH
\cite{Shen:2011eg} and \V \cite{Song:2011qa}. 
%
%
A straightforward extrapolation with fixed $\es$ overpredicts the LHC $v_2^\mathrm{ch}$ 
values by 10-15\%; a slight increase of $(\eta/s)_\mathrm{QGP}$ from 0.16 to 0.20 (for 
MC-KLN) gives better agreement with the ALICE data \cite{Aamodt:2010pa}. However, at 
LHC energies $v_2$ becomes sensitive to details of the initial shear stress profile 
\cite{Shen:2011eg}, and no firm conclusion can be drawn yet  whether the QGP turns 
more viscous (i.e. less strongly coupled) at higher temperatures. Furthermore, ALICE 
\cite{Floris:2011ru} has noted a discrepancy between the $\bar p/\pi^-$ ratio measured
in Pb+Pb collisions at the LHC and the value observed by STAR in Au+Au collisions
at RHIC. The latter has a strong influence on the value of the chemical decoupling
temperature implemented in the model. We use $\Tchem\eq165$\,MeV which nicely
fits the normalization of the proton spectra from STAR but overpredicts those from
PHENIX by a factor $\sim 1.5{-}2$ (left panel in Fig.~\ref{F2}). The $\bar p/\pi^-$ ratio
measured by ALICE at the LHC agrees with the PHENIX value measured at RHIC
(see Fig.~7 in \cite{Floris:2011ru}) but is smaller by a factor $\sim 1.5{-}2$ than what 
is implemented in the LHC predictions from Refs.~\cite{Shen:2011eg,Song:2011qa}.
Correspondingly, our predictions of the $\bar p$ spectra for Pb+Pb@LHC 
\cite{Shen:2011eg} overpredict the measured spectra by this factor \cite{Floris:2011ru}. 
Reducing the $\bar p/\pi^-$ ratio to the measured value will reduce the charged hadron 
elliptic flow \cite{Hirano:2002ds}. To to the larger radial flow, this reductions is
stronger at the LHC than at RHIC. This may account for the $\sim15\%$ overprediction of 
$v_2^\mathrm{ch}$ for $\es\eq0.16$ at the LHC seen in the left panel of Fig.~\ref{F3}.

The right panel of this figure shows that, at fixed $p_T{\,<\,}1$\,GeV, $v_2(p_T)$ 
increases from RHIC to LHC for pions but decreases for all heavier hadrons. The 
similarity at RHIC and LHC of  $v_2^\mathrm{ch}(p_T)$ for the sum of all charged 
hadrons noted in Ref.~\cite{Aamodt:2010pa} thus appears accidental. As a result
of this shift of the elliptic flow to larger $p_T$ for heavier particles, which is caused by
the stronger radial flow at the LHC, the mass-splitting between the $v_2(p_T)$ curves
for different mass hadrons grows from RHIC to LHC. This predicted growth has been
confirmed by ALICE (see Fig.~6 in \cite{Collaboration:2011yba}).

\begin{figure}[h!]
  \includegraphics[width=\linewidth,height=8.8cm,clip=,angle=0]{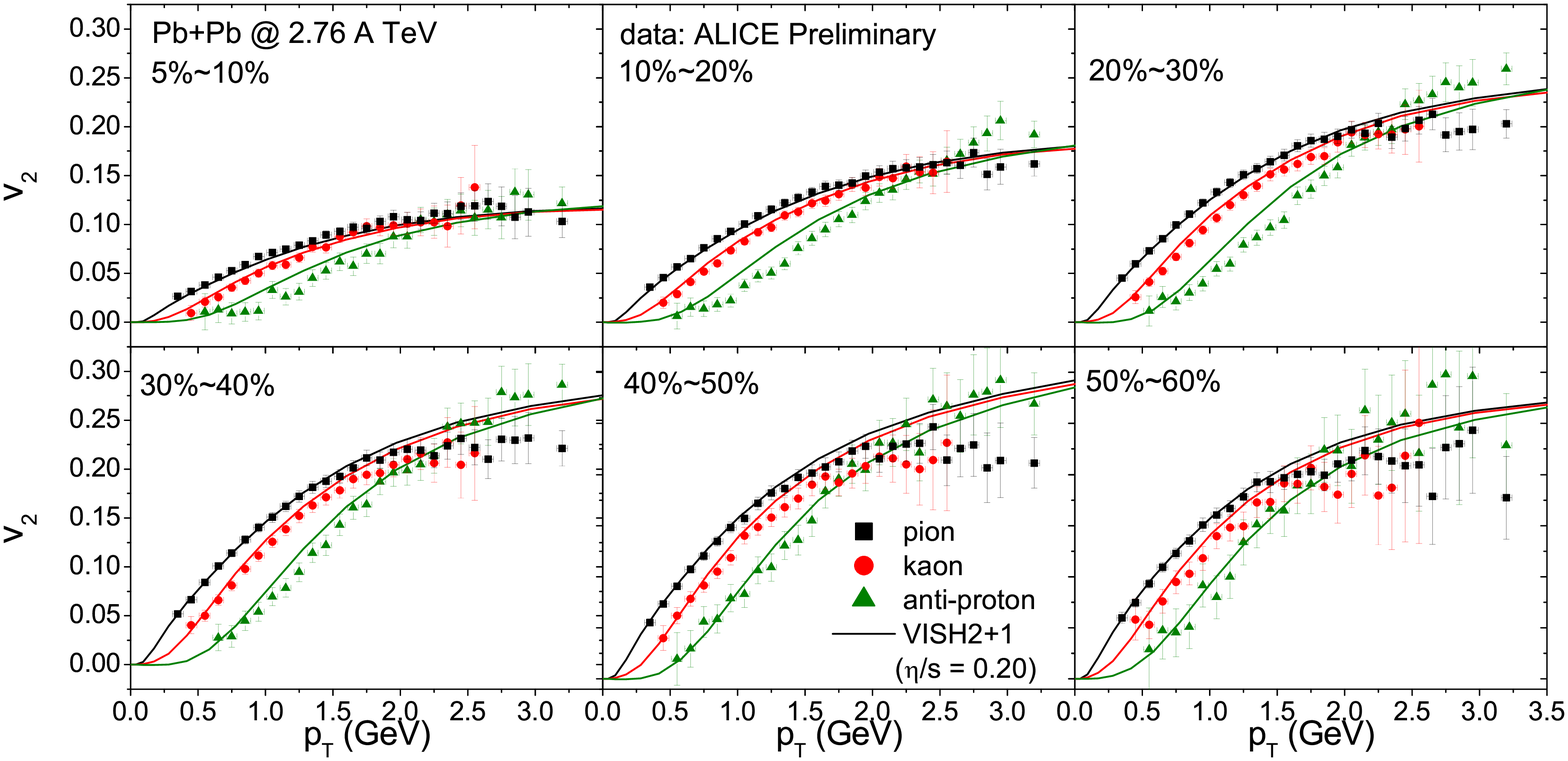}
\caption{\label{F4} (Color online)   Differential elliptic flow, $v_2\{2\}(p_T)$, for 
  pions, kaons and anti-protons from 2.76\,$A$\,TeV Pb+Pb collisions at different centralities,
  as measured by ALICE (preliminary data reported at {\it Quark Matter 2011}
  \cite{Collaboration:2011yba, Krzewicki:2011ee}), compared
  with VISH2{+}1 calculations using MC-KLN initial conditions with $\eta/s\eq0.2$
  \cite{Shen:2011eg}. 
       }
\end{figure}

As mentioned in footnote 2, the purely hydrodynamic simulations based on \VISH
with constant $\eta/s\eq0.2$ fail to correctly reproduce the centrality dependence
of the proton elliptic flow $v_2^p(p_T)$. Especially in central collisions, $v_2^p(p_T)$
is overpredicted at small $p_T$ (see Fig.~2 in \cite{Krzewicki:2011ee}), {\it i.e.}
the radial flow pushing the elliptic flow to higher $p_T$ (and thereby reducing
$v_2(p_T)$ at low $p_T$) as generated by the model is not strong enough in central 
collisions. The same problem is seen in Fig.~\ref{F4} for the extrapolation of \VISH
to LHC energies: whereas the pion and kaon elliptic flows at $p_T{\,<\,}1.5$\,GeV/$c$ 
are well described at all collision centralities, there appears to be a lack of radial
flow in central and semi-central collisions such that the proton $v_2$ is not pushed
towards larger $p_T$ as strongly as seen in the data. This problem disappears in 
the more peripheral bins, indicating an incorrect centrality dependence of the
balance between radial and elliptic flow in the \VISH model.    

%
%

As was the case at RHIC energies, this problem is removed when replacing the
hydrodynamic description of the late hadronic phase with low $\eta/s\eq0.2$ by
a microscopic kinetic description using \V (see Fig.~\ref{F5}). The calculations shown
in Fig.~\ref{F5} were done primarily to understand systematic differences between
the predictions for spectra and elliptic flow from \VISH and {\small VISHNU}. For
this reason they were performed with the same value $\eta/s\eq0.2$ in the QGP phase,
even though the \V calculations for identified hadrons from Au+Au at RHIC 
\cite{Song:2011hk} had shown a slight preference for the smaller value $\es\eq0.16$.
Fig.~\ref{F5} shows that the problem with the lack of radial flow in central collisions
that was seen in Fig.~\ref{F4} has been resolved: for the 5\%-10\% and 10\%-20\% 
centrality bins, \V describes the differential $v_2$ up to $p_T\eq2$\,GeV/$c$ almost 
perfectly, for all three particle species. Looking closely, one observes a slight
(6\%) underprediction of $v_2(p_T)$ {\it for all three particle species}. This underprediction
gets stronger in more peripheral collisions (reaching 9\% at 50\%-60\% centrality) -- a 
clear sign that we have slightly overestimated $\es$. (Since, for fixed $\eta/s$, viscous 
effects increase in inverse proportion to the fireball size \cite{Song:2007fn}, an 
overestimate of $\eta/s$ will lead to an underprediction of $v_2(p_T)$ that {\it grows} 
with the impact parameter of the collisions.)  We are confident that, after reducing $\es$ 
to 0.16, the data will be well described {\it at all collision centralities}. Corresponding 
simulations are in progress.

\section{Conclusions}
\label{sec4}

The hybrid model {\small VISHNU}, which describes the evolution of the dense and 
strongly coupled  quark-gluon plasma phase macroscopically using viscous fluid 
dynamics and that of the dilute late hadronic rescattering and freeze-out stage 
microscopically using a kinetic approach, provides a comprehensive quantitative 
description of the bulk matter created in relativistic heavy-ion collisions at RHIC and LHC.
Transverse momentum spectra and elliptic flow of soft charged hadrons, pions, kaons, and 
protons are well reproduced at all collision centralities, with a QGP shear viscosity 
$\es\eq\frac{2}{4\pi}\eq0.16$ if MC-KLN initial conditions are used. So far the data yield 
no  evidence for a change of $\es$ between RHIC and LHC that would reflect the different temperature ranges probed. Overall, the QGP liquid created in heavy-ion collisions at the LHC appears to be as strongly coupled as at RHIC energies. 

%
\begin{figure}[h!]
 \includegraphics[height=5.62cm,clip=,angle=270]{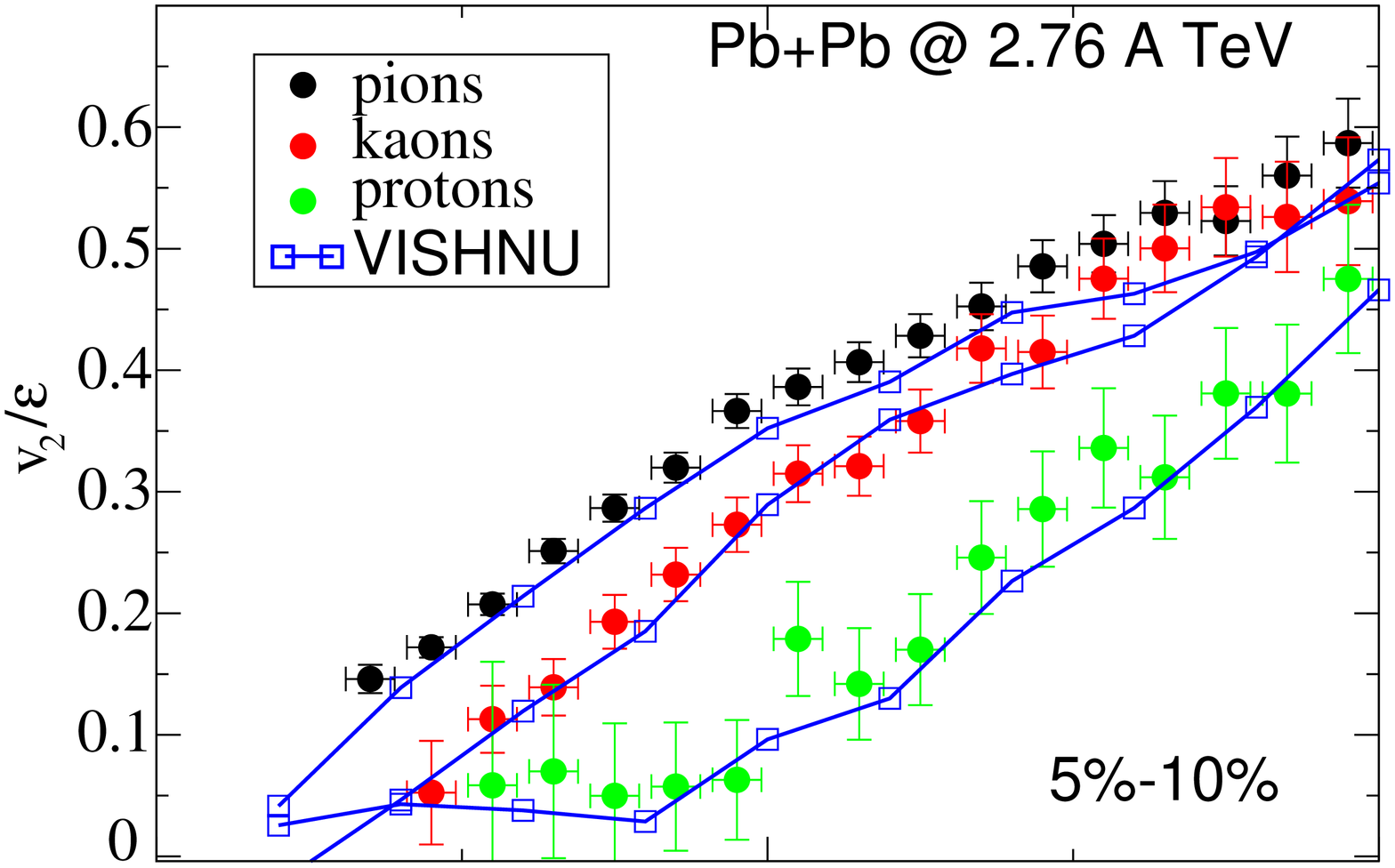}
 \hspace*{-1.5mm}
 \includegraphics[height=5cm,clip=,angle=270]{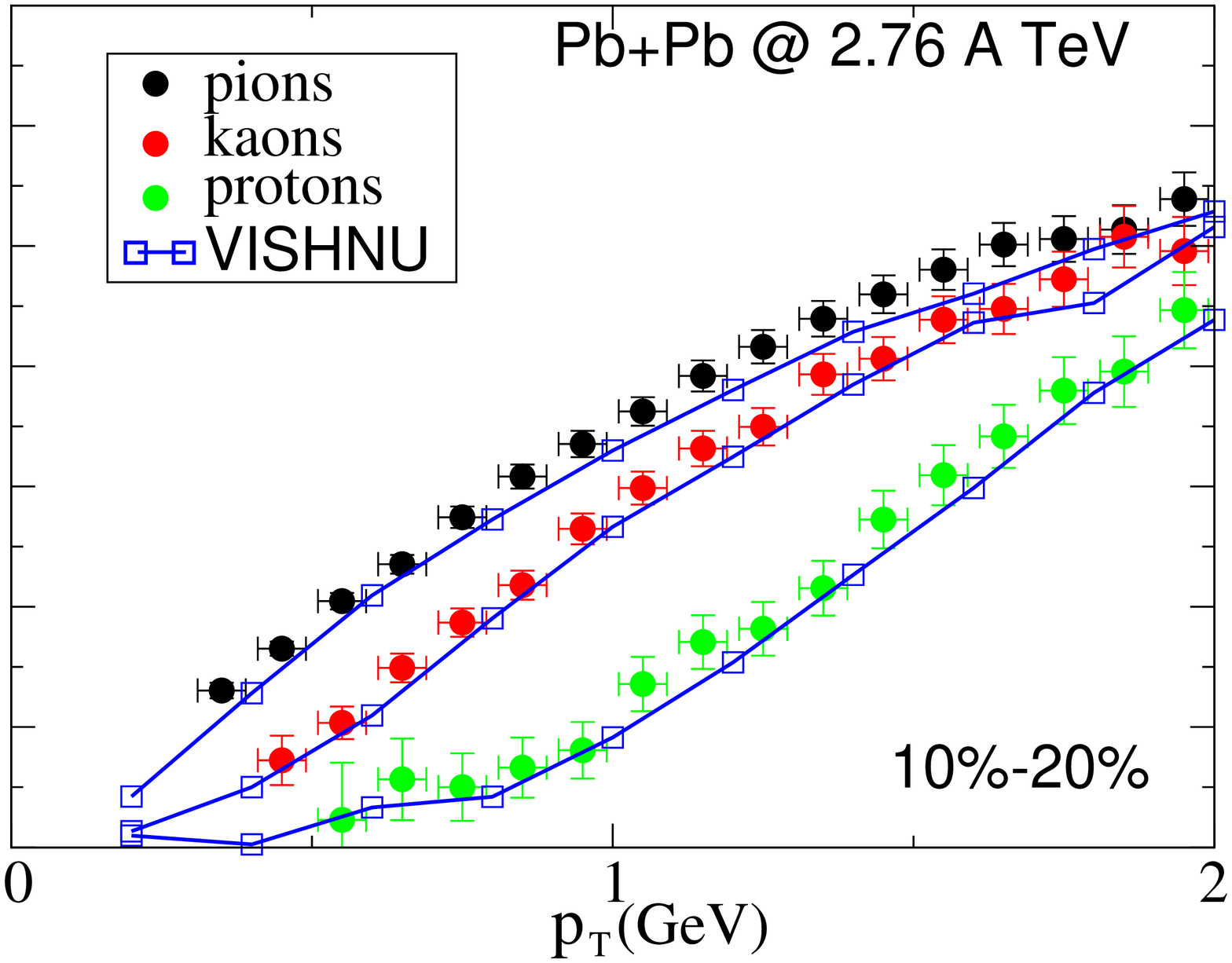}
 \hspace*{-1.5mm}
 \includegraphics[height=5cm,clip=,angle=270]{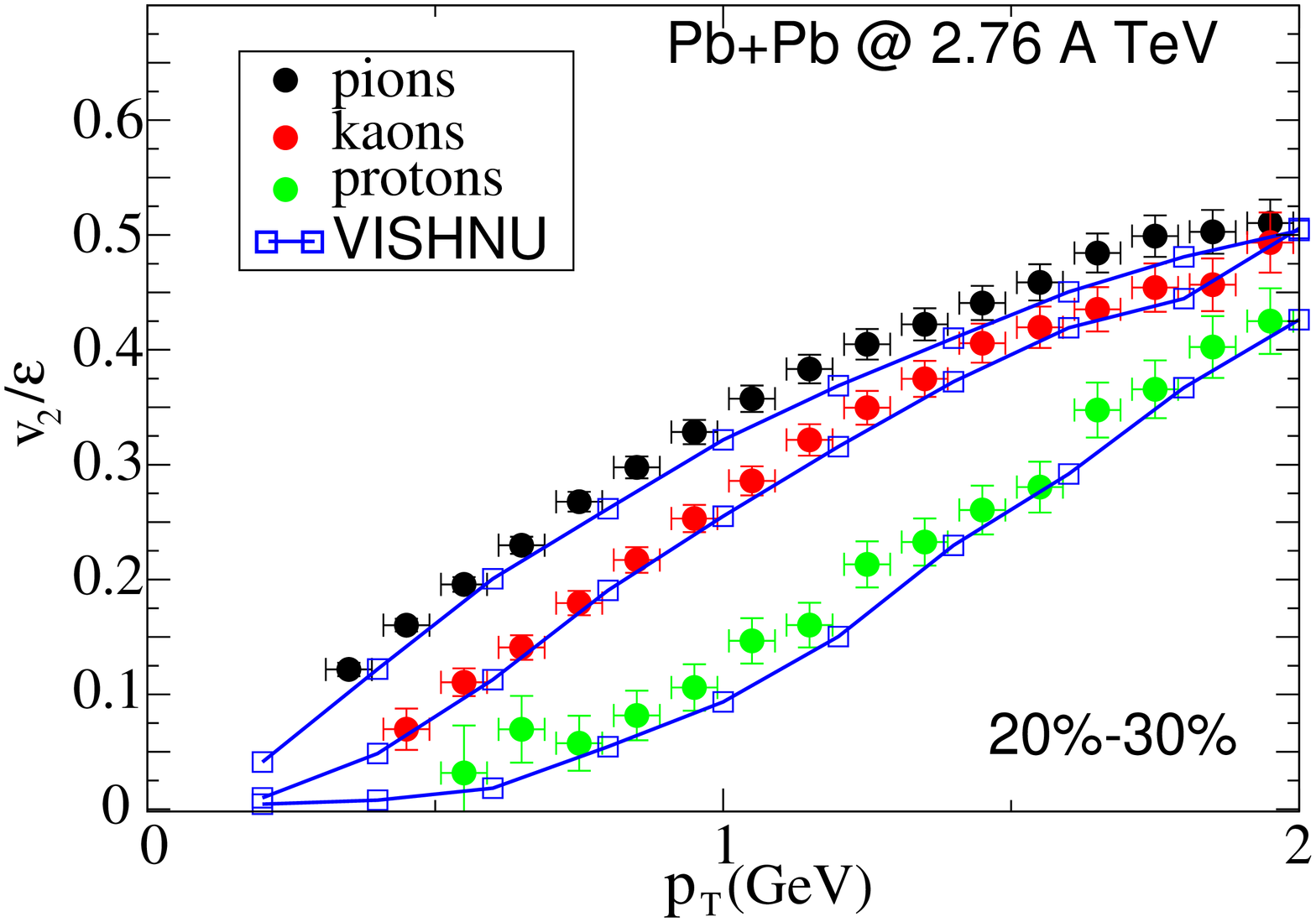}
\end{figure} 
\vspace*{-3.8mm}
\begin{figure}[h!]
 \includegraphics[height=5.62cm,clip=,angle=270]{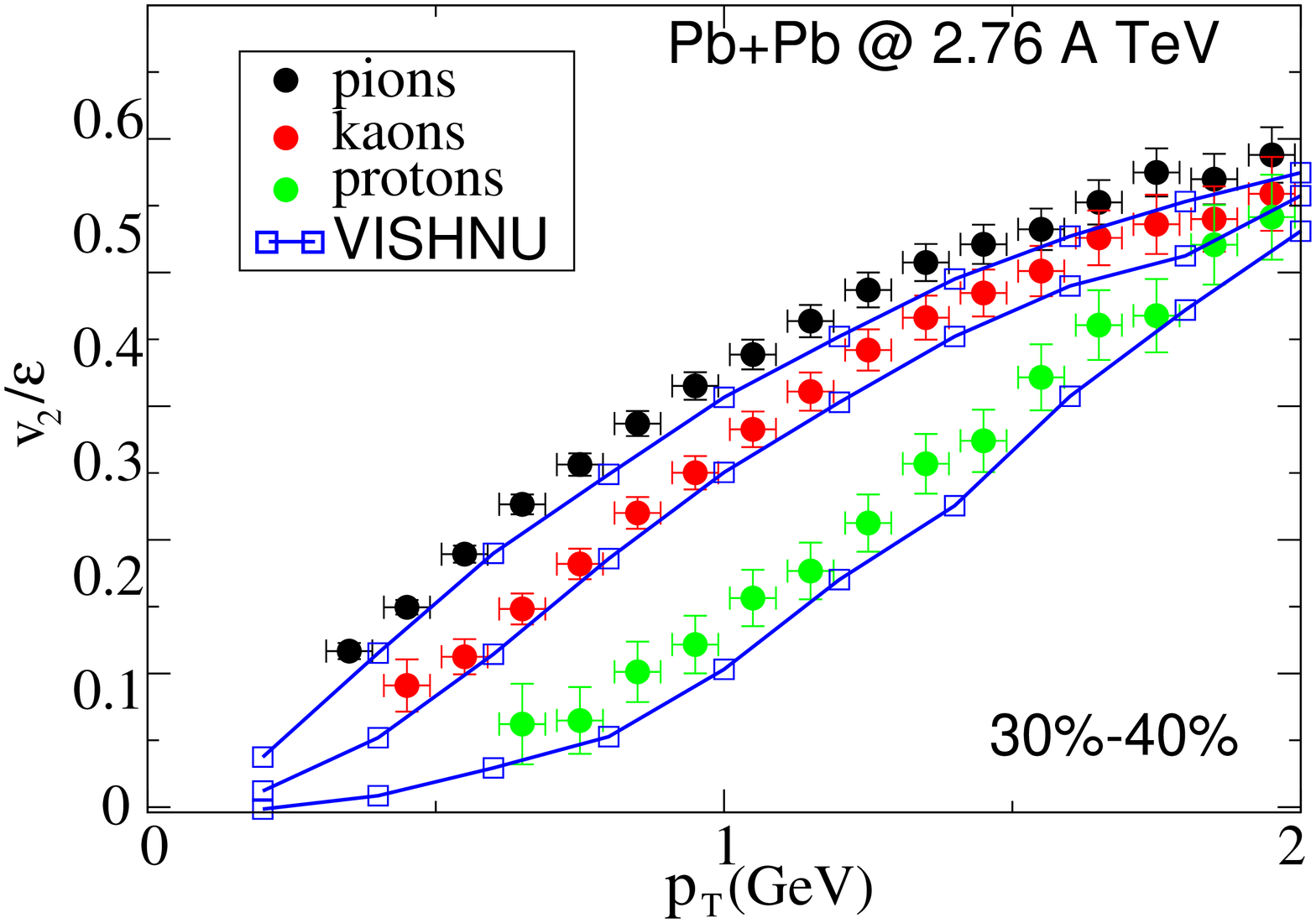}
 \hspace*{-1.5mm}
 \includegraphics[height=5cm,clip=,angle=270]{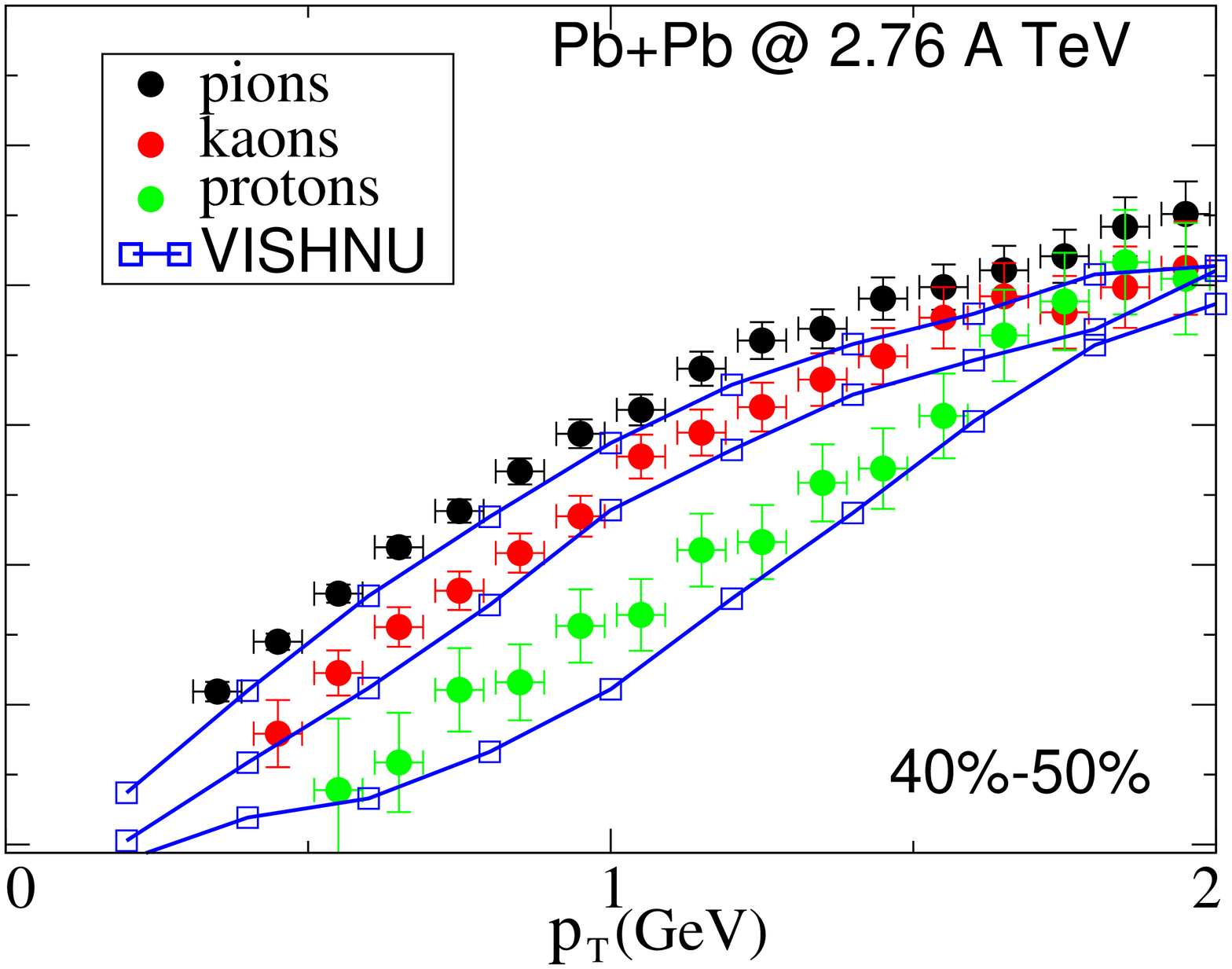}
 \hspace*{-1.5mm}
 \includegraphics[height=5cm,clip=,angle=270]{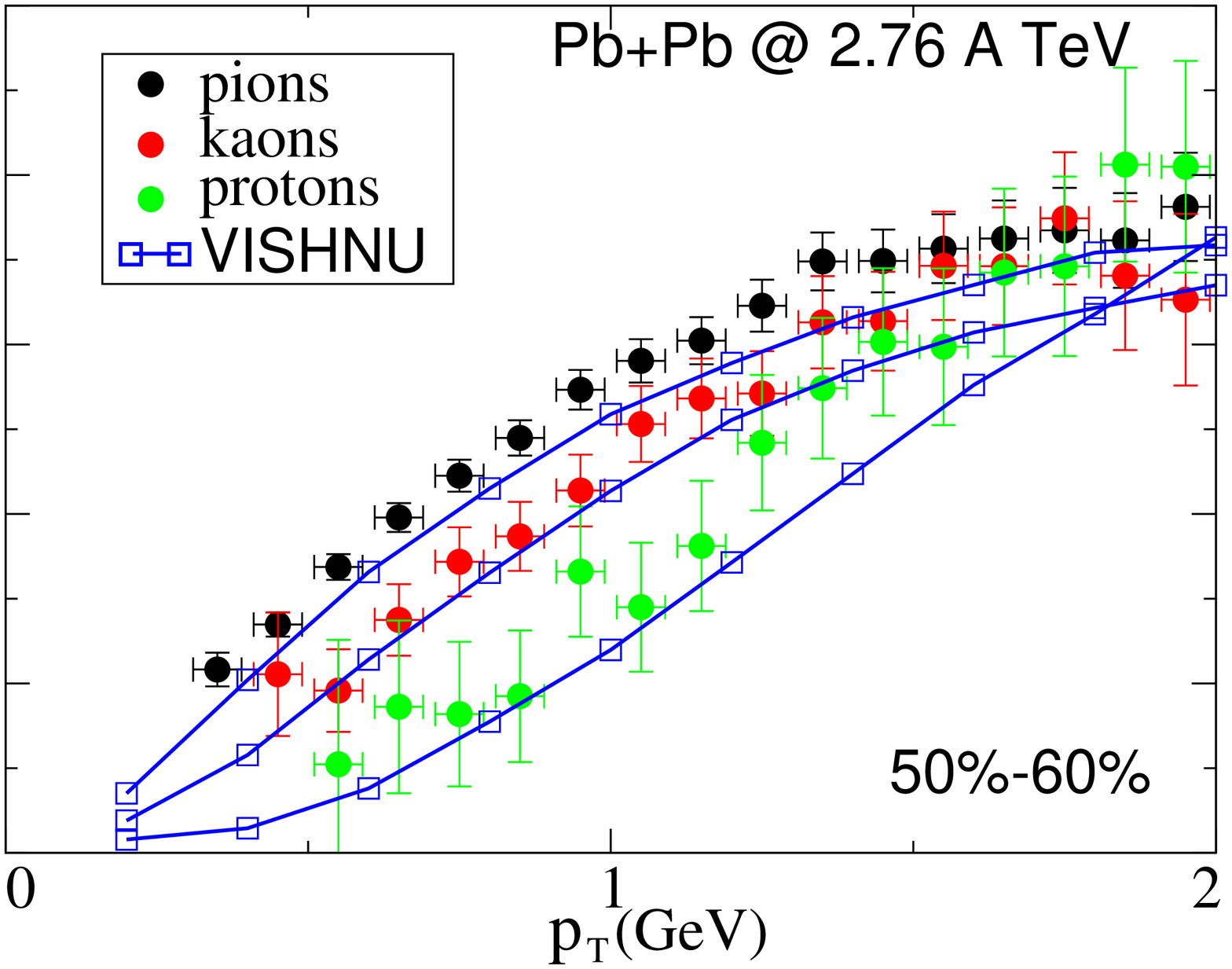}
\caption{\label{F5} (Color online)  Same preliminary data from ALICE  
   \cite{Collaboration:2011yba, Krzewicki:2011ee} as in Fig.~\ref{F4}, but now
   compared with VISHNU calculations with $\es\eq0.2$, using the same MC-KLN
   initial conditions as in Fig.~\ref{F3}. Shown is the eccentricity-scaled elliptic flow,
   {\it i.e.} $v_2\{2\}/\ecc_x\{2\}$ for the experimental data and $\La v_2\Ra/\La\ecc_x\Ra$
   for the theoretical curves.
   }
\end{figure}
%

\bigskip


\noindent
{\bf Acknowledgments:} This work was supported by the U.S.\ Department of Energy under grants 
No. DE-AC02-05CH11231, \rm{DE-SC0004286} and (within the framework of the JET
Collaboration) No. \rm{DE-SC0004104}. Extensive computing resources provided by the Ohio
Supercomputing Center are gratefully acknowledged.



\bibliographystyle{aipproc}   

\begin{thebibliography}{99}

\bibitem{Heinz:2005zg}
  U. Heinz, 
  {\it Preprint arXiv:nucl-th/0512051}.

\bibitem{Hirano:2007xd}
  T. Hirano, U. Heinz, D. Kharzeev, R. Lacey and Y. Nara, 
  {\it J.\ Phys.} G {\bf 34}, S879 (2007).

\bibitem{Hirano:2005xf}
  T. Hirano, U. Heinz, D. Kharzeev, R. Lacey and Y. Nara,
  {\it Phys.\ Lett.} {\bf B636}, 299 (2006).

\bibitem{Bass:1998ca}
  S. A. Bass {\it et al.},
  {\it Prog.\ Part.\ Nucl.\ Phys.} {\bf 41}, 255 (1998).

\bibitem{Song:2010aq}
  H. Song, S. A. Bass and U. Heinz,
  {\it Phys.\ Rev.} C {\bf 83}, 024912 (2011).

\bibitem{Song:2007fn}
  H.~Song and U.~Heinz,
  {\it Phys.\ Lett.}\  {\bf B658}, 279-283 (2008);
  {\it Phys.\ Rev.}\  C {\bf 77}, 064901 (2008);
  and 
  {\it Phys.\ Rev.}\ C {\bf 78}, 024902 (2008).

\bibitem{Song:2010mg}
  H. Song, S. A. Bass, U. Heinz, T. Hirano and C. Shen,
  {\it Phys.\ Rev.\ Lett.}  {\bf 106}, 192301 (2011).

\bibitem{Ollitrault:2009ie}
  J.-Y. Ollitrault, A. M. Poskanzer and S. A. Voloshin 
  {\it Phys.\ Rev.} C {\bf 80}, 014904 (2009).
    
\bibitem{Shen:2011zc}
  C.~Shen, S.~A.~Bass, T.~Hirano, P.~Huovinen, Z.~Qiu, H.~Song, U.~Heinz,
  in {\it Quark Matter 2011}, {\it J. Phys.} G, in press ({\it Preprint
  arXiv:1106.6350 [nucl-th]}).

\bibitem{Qiu:2011fi}
  Z.~Qiu and U.~Heinz,
 these proceedings ({\it Preprint arXiv:1108.1714 [nucl-th]}).
 
 \bibitem{:2011vk}
   K. Aamodt {\it et al.}  [ALICE Collaboration],
  {\it Phys.\ Rev.\ Lett.}\  {\bf 107}, 032301 (2011).

\bibitem{Adare:2011tg}
  A.~Adare {\it et al.}  [PHENIX Collaboration],
  {\it Preprint arXiv:1105.3928 [nucl-ex]}.
 
\bibitem{Qiu:2011iv}
  Z.~Qiu and U.~Heinz,
  {\it Phys.\ Rev.}\ C {\bf 84}, 024911 (2011). 

\bibitem{Song:2011hk}
  H. Song, S. A. Bass, U. Heinz, T. Hirano and C. Shen,
  {\it Phys.\ Rev.} C {\bf 83}, 054910 (2011).

\bibitem{Shen:2011eg}
  C.~Shen, U.~Heinz, P.~Huovinen and  H.~Song,
  {\it Preprint arXiv:1105.3226 [nucl-th]}.

\bibitem{Song:2011qa}
  H.~Song, S.~A.~Bass and U.~Heinz,
  {\it Phys.\ Rev.}\  C {\bf 83}, 054912 (2011).

\bibitem{Aamodt:2010pa}
  K. Aamodt {\it et al.}  [ALICE Collaboration],
  {\it Phys.\ Rev.\ Lett.} {\bf 105}, 252302 (2010).

\bibitem{Floris:2011ru}
  M.~Floris {\it et al.} [ALICE Collaboration],
  in {\it Quark Matter 2011}, {\it J. Phys.} G, in press 
  ({\it Preprint arXiv:1108.3257 [hep-ex]}).

\bibitem{Hirano:2002ds}
  T.~Hirano and K.~Tsuda,
  {\it Phys.\ Rev.}\ C {\bf 66}, 054905 (2002);
  P.~F.~Kolb and R.~Rapp,
  {\it ibid.} {\bf 67}, 044903 (2003);
  P.~Huovinen,
  {\it Eur.\ Phys.\ J.}\  {\bf A37}, 121 (2008).

\bibitem{Collaboration:2011yba}
  R.~Snellings {\it et al.} [ALICE Collaboration],
  in {\it Quark Matter 2011}, {\it J. Phys.} G, in press ({\it Preprint arXiv:1106.6284 [nucl-ex]}).

\bibitem{Krzewicki:2011ee}
  M.~Krzewicki {\it et al.} [ALICE Collaboration],
  in {\it Quark Matter 2011}, {\it J. Phys.} G, in press ({\it Preprint arXiv:1107.0080 [nucl-ex]}).

\end{thebibliography}

\end{document}